\documentclass{acm_proc_article-sp}
\usepackage{amssymb}
\usepackage{enumitem}
\setenumerate{nolistsep}
\setitemize{nolistsep}
\usepackage{dsfont}
\usepackage[usenames, dvipsnames]{color}
\usepackage{tikz}
\usetikzlibrary{automata,positioning}
\usepackage{subfig}
\usepackage[colorlinks=true,
		linkcolor=MidnightBlue,
		citecolor=Mahogany,
		urlcolor=ForestGreen,
		pdfauthor={Atreyee Kundu and Debasish Chatterjee},
		pdftitle={Global asymptotic stability of switched linear systems},
		pdfkeywords={switched systems, asymptotic stability},
		pdfsubject={Technical Report},
		plainpages=false]{hyperref}

\usepackage{mathtools}
\allowdisplaybreaks
\newtheorem{theorem}{Theorem}
\newdef{definition}{Definition}
\newdef{fact}{Fact}
\newdef{assumption}{Assumption}
\newdef{prop}{Proposition}
\newdef{remark}{Remark}
\newdef{example}{Example}
\newdef{problem}{Problem}
\newcommand{\norm}[1]{\left\lVert{#1}\right\rVert}
\newcommand{\abs}[1]{\left\lvert{#1}\right\rvert}

\newcommand{\Let}{\coloneqq}
\newcommand{\teL}{\eqqcolon}
\newcommand{\nn}{\nonumber}

\newcommand{\pmat}[1]{\begin{pmatrix}#1\end{pmatrix}}

\renewcommand{\geq}{\geqslant}
\renewcommand{\ge}{\geqslant}
\renewcommand{\leq}{\leqslant}
\renewcommand{\le}{\leqslant}
\newcommand{\R}{\mathds{R}}
\newcommand{\N}{\mathds{N}}
\renewcommand{\P}{\mathcal{P}}
\newcommand{\ls}{\limsup}
\newcommand{\li}{\liminf}

\newcommand{\Ntsigma}{N_{t}^{\sigma}} %number of switchings till t
\newcommand{\ktol}{\#{\{k\rightarrow\ell\}_{t}}} %no. of transitions from k to l
\newcommand{\indicator}{\mathds{1}_{\{j\}}(\sigma(\tau_{i}))} %indicator function
\DeclareMathOperator{\minimize}{minimize}
\DeclareMathOperator{\sbjto}{subject\;to}
\begin{document}

%====================================================================
% Title begins
\title{Stabilizing discrete-time switched linear systems}

\numberofauthors{2}
\author{
\alignauthor Atreyee Kundu\\
       \affaddr{Systems \& Control Engineering}\\
       \affaddr{Indian Institute of Technology Bombay}\\
       \affaddr{Mumbai 400076, India}\\
       \email{atreyee@sc.iitb.ac.in}
\and
\alignauthor Debasish Chatterjee\\
       \affaddr{Systems \& Control Engineering}\\
       \affaddr{Indian Institute of Technology Bombay}\\
       \affaddr{Mumbai 400076, India}\\
       \email{dchatter@iitb.ac.in}
}

\maketitle
%==========================================================================

%===========================================================================
\begin{abstract}
    This article deals with stabilizing discrete-time switched linear systems. Our contributions are threefold: Firstly, given a family of linear systems possibly containing unstable dynamics, we propose a large class of switching signals that stabilize a switched system generated by the switching signal and the given family of systems. Secondly, given a switched system, a sufficient condition for the existence of the proposed switching signal is derived by expressing the switching signal as an infinite walk on a directed graph representing the switched system. Thirdly, given a family of linear systems, we propose an algorithmic technique to design a switching signal for stabilizing the corresponding switched system.
\end{abstract}
%=============================================================================

%\terms{}

%===============================================================================
%===============================================================================

%================================================================================
\section{Introduction}
\label{s:intro}
%================================================================================
    A \emph{switched system} \cite[\S1.1.2]{Liberzon} comprises of two components --- a finite family of systems and a \emph{swicthing signal} that selects an active subsystem from the family at every instant of time. Switched systems arise in a multitude of application areas such as networked systems, quantization, variable structure systems, etc; see e.g., \cite{Heemels_survey, Heemels_switched, Heemels_lecture, Shorten_review, Antsaklis_survey} and the references therein.

     Stability of switched systems has attracted considerable research attention over the past few decades, see \cite{Antsaklis_survey, Heemels_survey, Shorten_review} for detailed surveys. In this article we study stability of discrete-time switched linear systems under \emph{constrained switching} \cite[Chapter 3]{Liberzon} as a continuation of our study of switched systems initiated in \cite{abc}. To wit, we are concerned with identifying classes of switching signals for which the switched system under consideration is globally asymptotically stable, as well as algorithmic synthesis of stabilizing switching signals.

     Conditions for stability under constrained switching typically employ the idea of \emph{slow switching} vis-a-vis \emph{(average) dwell time} switching. Although these results were originally developed in the setting of continuous-time systems \cite[Chapter 3]{Liberzon}, \cite{HespanhaMorse}, they can be readily extended to the discrete-time setting, with the (average) dwell time expressed in terms of the number of time steps \cite{Antsaklis_survey}. Most of these results apply to switched systems with all stable subsystems.

	 In the presence of unstable dynamics in the family, slow switching alone is not sufficient to guarantee stability of the switched system. Additional conditions are required to ensure that the switched system does not spend too much time on the unstable subsystems; see e.g., \cite{Antsaklis_survey}. In \cite{Zhai002} the authors discuss a class of switching signals for global exponential stability of a switched system in which not all subsystems are Schur stable (or even when no subsystem is Schur stable) but the unstable subsystems form a stable combination. The characterization of the stabilizing switching signal involves a modified definition of average dwell time and the method of activating the Schur stable subsystems (if any) arbitrarily but activating the unstable subsystems depending on a pre-specified ratio. However, none of the above works ventured beyond switching signals with switching \emph{frequencies} that are faster than a constant.

    Keeping to the general tune of slow switching while extending its scope considerably, in this article we consider switched systems with unstable dynamics and design a class of stabilizing switching signals that transcends beyond the (average) dwell time regime. Moreover, given a switched system, we derive a sufficient condition for the existence of the proposed switching signal. This matter demands attention in the context of switched systems because given a switched system, a purely time-dependent stabilizing switching signal may be difficult to find. We also present an algorithmic approach to construct a switching signal that satisfies the proposed condition. More specifically, our contributions are:
    \begin{itemize}[label=\(\circ\), leftmargin=*]
        \item We propose a family of stabilizing switching signals such that:
        \begin{itemize}[label=$\triangleright$, leftmargin=*]
            \item The switching signals are more general than those having a certain average dwell time. In particular, switching rates that are faster than a constant are admissible.
            \item We specify \emph{only} certain asymptotic properties of the switching signal; there is, therefore plenty of flexibility insofar as the transient behaviour of the switching signals are concerned.
            \item We allow unstable dynamics in the family of systems. Although this is not the first instance that stability of a discrete-time switched linear system containing unstable dynamics in the family has been considered, to the best of our knowledge, this is the first instance when a stabilizing switching signal for a discrete-time switched linear system containing unstable subsystems, is proposed \emph{solely} in terms of asymptotic behaviour of the switching signal.
        \end{itemize}
        \item We develop conditions for existence of switching signals having the properties that we propose. These conditions are algorithmic in nature, and can be readily tested for a given switched system.
        \item We provide an algorithmic procedure for synthesis of switching signals that ensure stability of discrete-time switched linear systems.
    \end{itemize}
    The analysis and the synthesis carried out in this article can be extended in a standard fashion to include the case of switched nonlinear systems; we do not strive for maximum generality in terms of applicability of the ideas presented here.

    The remainder of this article is organized as follows: In \S\ref{s:prelim} we formulate the problem under consideration and elaborate on the properties of the family of systems and the switching signal that we employ in our analysis. Our main results are stated in \S\ref{s:mainres}, and we illustrate our results with the aid of a numerical example in \S\ref{s:ex}. We conclude in \S\ref{s:concln} with a brief summary of some of the natural future directions. The proofs of all facts and claims are presented in a consolidated fashion in \S\ref{s:allproofs}.

    {\bf Notation.} $\N = \{1,2,\cdots\}$ is the set of natural numbers, $\N_{0} = \{0\}\cup\N$, and $\R$ is the set of real numbers. We denote by $k_{1}:k_{2}$ the set $\{n\in\N_{0}:k_{1}\le n\le k_{2}\}$. We let $\mathds{1}_{A}(\cdot)$ denote the indicator function of a set $A$, and $I_{d\times d}$ denote the $d$-dimensional identity matrix. Let $\norm{\cdot}$ be the standard $2$-norm and let $^\top$ denote the transpose operation.
%=====================================================================

%=====================================================================
\section{Preliminaries}
\label{s:prelim}
%======================================================================
    We consider a discrete-time \emph{switched linear system}
    \begin{align}
    \label{e:swsys}
        x(t+1) = A_{\sigma(t)}x(t), \:x(0)\:\text{given}, \:t\in\N_{0},
    \end{align}
    generated by the following two components:
    \begin{itemize}[label=\(\circ\), leftmargin=*]
        \item a family of systems
            \begin{align}
            \label{e:family}
                x(t+1) = A_{i}x(t),\:x(0)\:\text{given},\:i\in\P,\:t\in\N_{0},
            \end{align}
            where $x(t)\in\R^{d}$ is the vector of system states at time $t$, $\P$ is a finite index set, and $A_{i}\in\R^{d\times d}$ are known constant matrices.
        \item a \emph{switching signal} $\sigma:\N_{0}\:\to\P$ specifying at every time $t$, the index of the active subsystem from the family \eqref{e:family}.
    \end{itemize}

    We assume that for each $i\in\P$ the matrix $A_{i}\in\R^{d\times d}$ has full rank. Consequently, $0\in\R^{d}$ is the unique equilibrium point for each system in the family \eqref{e:family}.
    Let $0 \teL \tau_{0}<\tau_{1}<\cdots$ be the switching instants, i.e., those positive integers at which $\sigma$ changes values. For $t > 0$ let $\Ntsigma$ denote the number of switches (before and including) $t$. The solution $(x(t))_{t\in\N_{0}}$ to the switched system \eqref{e:swsys} corresponding to a switching signal $\sigma$ is given by
    \begin{align}
    \label{e:sol}
        x(t) = A_{\sigma(\tau_{\Ntsigma})}^{t-\tau_{\Ntsigma}}A_{\sigma(\tau_{\Ntsigma-1})}^{\tau_{\Ntsigma}-\tau_{\Ntsigma-1}}\cdots
        A_{\sigma(0)}^{\tau_{1}-\tau_{0}}x(0), \quad t\in\N_{0},
    \end{align}
    where the dependence of $(x(t))_{t\in\N_{0}}$ on $\sigma$ has been suppressed.

    In this article we characterize a class of switching signals under which \eqref{e:swsys} is globally asymptotically stable (GAS). By definition,
    \begin{definition}
    \label{d:gas}
        The switched system \eqref{d:gas} is \emph{globally asymptotically stable} (GAS) for a given switching signal $\sigma$ if \eqref{e:swsys} is
        \begin{itemize}[label=\(\circ\), leftmargin=*]
            \item Lyapunov stable, and
            \item globally asymptotically convergent, i.e., irrespective of the initial condition $x(0)$, $x(t) \rightarrow 0$ as $t\rightarrow +\infty$.
        \end{itemize}
    \end{definition}

    Prior to presenting our main result, we identify and derive key properties of the family of systems \eqref{e:family} and the switching signal $\sigma$, that will be used in our analysis.

%==============================================================
    \subsection{Properties of the family \eqref{e:family}}
    \label{ss:familyprop}
%===============================================================
     Given the family of systems \eqref{e:family}, the following fact captures the quantitative measure of (in)stability of each system.
     \begin{fact}
     \label{fact:key}
       For each $i\in\P$ there exists a pair $(P_{i},\lambda_{i})$, where $P_{i}\in\R^{d\times d}$ is a symmetric and positive definite matrix, and
       \begin{itemize}[label=\(\circ\), leftmargin=*]
        \item if $A_{i}$ is asymptotically stable, then $0<\lambda_{i}<1$;
        \item if $A_{i}$ is marginally stable, then $\lambda_{i}=1$; \footnote{Marginally stable systems are those systems in \eqref{e:family} that are Lyapunov stable but not asymptotically stable.}
         \item if $A_{i}$ is unstable, then $\lambda_{i} > 1$;
        \end{itemize}
        such that, with
        \begin{equation}
        \label{e:Lyapdefn}
            \R^{d}\ni\xi\longmapsto V_{i}(\xi)\Let\langle P_{i}\xi, \xi\rangle\in[0,+\infty[,
        \end{equation}
        we have
        \begin{equation}
        \label{e:Lyapprop}
            V_{i}(\gamma_i(t+1))\leq \lambda_{i}V_{i}(\gamma_i(t)),\quad t\in\N_{0},
        \end{equation}
        and $\gamma_i(\cdot)$ solves the $i$-th recursion in \eqref{e:family}, $i\in\P$.
    \end{fact}
    The functions $V_{i}$, $i\in\P$, defined in \eqref{e:Lyapdefn} will be called Lyapunov-like functions in the sequel. Even though Fact \ref{fact:key} is a folklore result, in \S\ref{s:proofs} we sketch a proof for completeness.

    We observe that for all $i,j\in\P$ the respective Lyapunov-like functions are related as follows:
    \begin{fact}
    \label{fact:muij}
        There exists $\mu_{ij} > 0$ such that
        \begin{align}
        \label{e:Lyapreln}
            V_{j}(\xi) \leq \mu_{ij} V_{i}(\xi) \quad \text{for all} \quad \xi\in\R^{d},
        \end{align}
        whenever the switching from system $i$ to system $j$ is admissible.
    \end{fact}
    The assumption that there exists $\mu\ge 1$ such that $V_{j}(\xi) \leq \mu V_{i}(\xi)$ for all $i,j\in\P$ and $\xi\in\R^{d}$ is standard \cite{Zhai002}. Clearly, the preceding inequality is a special case of \eqref{e:Lyapreln}. A \emph{tight} estimate of the numbers $\mu_{ij}$ may be given as follows; we present a short proof of this estimate in \S\ref{s:proofs}.
    \begin{prop}
    \label{p:muprop}
        Let the Lyapunov-like functions be defined as in \eqref{e:Lyapdefn} with each $P_{i}$ symmetric and positive definite, $i\in\P$. Then the smallest constant $\mu_{ij}$ in \eqref{e:Lyapreln} is given by
        \begin{align}
        \label{e:muij}
            \mu_{ij} = \lambda_{\max}(P_{j}P_{i}^{-1}), \quad i,j\in\P,
        \end{align}
        where for a matrix $M\in\R^{n\times n}$ having real spectrum, $\lambda_{\max}(M)$ denotes its maximal eigenvalue.
    \end{prop}

%============================================================
    \subsection{Properties of the switching signal}
    \label{ss:swprop}
%============================================================
    Recall that for a switching signal $\sigma$, $0 \teL \tau_{0} < \tau_{1} < \cdots < \tau_{\Ntsigma}$ are the switching instants before (and including) $t > 0$. We define the \emph{i-th holding time} of a switching signal $\sigma$ to be
    \begin{align}
    \label{e:holdtime}
        S_{i+1} \Let \tau_{i+1} - \tau_{i}, \quad i = 0,1,\cdots,
    \end{align}
    where $\tau_{i}$ and $\tau_{i+1}$ denote two consecutive switching instants. We let
    \begin{align}
    \label{e:swfreq}
        \nu(t) \Let \frac{\Ntsigma}{t}, \quad t > 0,
    \end{align}
    denote the \emph{switching frequency} of $\sigma$ at time $t$.

    We associate a directed graph $G(\P,E(\P))$ with the given switched system \eqref{e:swsys} in the following fashion:\footnote{A directed graph is a set of nodes connected by edges, where each edge has a direction associated to it. Directed graphs have appeared before in the switched systems literature in \cite{Mancilladigraph,graph_dwell}.}  Let the finite index set $\P$ denote the set of vertices of the directed graph $G$, and let the set of edges $E(\P)$ of $G$ contain a directed edge from $i$ to $j$, $i,j\in\P$, whenever a switching from system $i$ to system $j$ is admissible. If it is admissible to dwell on a system $i\in\P$ for at least two consecutive time steps, the vertex $i$ has a self-loop.\footnote{A self-loop is an edge that connects a vertex to itself.} For example:
    \begin{example}
        Given a family of systems with $\P = \{1,2,3\}$. Let a switching from system $1$ to any other system and from system $2$ to system $3$ be admissible. Also, let it be admissible to dwell on systems $2$ and $3$ for at least two consecutive time steps. The corresponding directed graph representation is shown below:

    \begin{center}
        \begin{tikzpicture}[every path/.style={>=latex},every node/.style={auto}]
            \node[state]            (a) at (-1.5,0)  { 1 };
            \node[state]            (b) at (1.5,0)  { 2 };
            \node[state]            (c) at (0,-1.5) { 3 };

            \draw[->] (a) edge (b);
            \draw[->] (a) edge (c);
            \draw[->] (b) edge (c);
            \draw[->] (b) edge[loop right] (b);
            \draw[->] (c) edge[loop below] (c);
        \end{tikzpicture}
        \end{center}

        For the directed graph in figure
        \[
            E(\P) = \{(1,2),(1,3),(2,2),(2,3),(3,3)\},
        \]
        where the edges $(1,2), (1,3)$, and $(2,3)$ correspond to the admissible switches, and the self-loops $(2,2)$ and $(3,3)$ depict the fact that it is admissible to dwell on systems $2$ and $3$ for at least two consecutive time steps.
    \end{example}

    Recall \cite[p.4]{Bollobas} that a \emph{walk} $W$ on a directed graph $G(V,E)$ is an alternating sequence of vertices and edges, say $x_{0},e_{1},x_{1},$\\$e_{2},\cdots,e_{\ell},x_{\ell}$, where $x_{i}\in V$, $e_{i}= (x_{i-1},x_{i})\in E$, $0 < i \le \ell$. The length of a walk is its number of edges, counting repetitions, e.g., in the above case the length of the walk $W$ is $\ell$. In the sequel by the term \emph{infinite walk} we mean a walk of infinite length, i.e., it has infinitely many edges.

    \begin{fact}
    \label{fact:walk}
        The set of switching signals $\sigma:\:\N_{0}\to\P$ and the set of infinite walks on $G(\P,E(\P))$ (defined as above) are in bijective correspondence.
    \end{fact}
    A proof of Fact \ref{fact:walk} is sketched in \S\ref{s:proofs}. Fix $t > 0$. For a switching signal $\sigma$ and each pair $(k,\ell)\in E(\P)$ we let
    \begin{align}
    \label{e:rhokl}
        \rho_{k\ell}(t) \Let \ktol
    \end{align}
    denote the number of transitions from vertex (system) $k$ to vertex (system) $\ell$ made by $\sigma$ before (and including) $t$, and for each $j\in\P$ we define
    \begin{align}
    \label{e:kappaj}
        \kappa_{j}(t) \Let \#\{j\}_{t} = \sum_{\displaystyle{{\substack{i:\sigma(\tau_{i})=j\\i=0,1,\cdots,\Ntsigma}}}}S_{i+1}
    \end{align}
    to be the number of times the vertex (system) $j$ is activated before (and including) $t$ by $\sigma$. Here $S_{i+1}$ is as defined in \eqref{e:holdtime}.

    In the latter half of \S\ref{s:mainres} we present some graph theoretic arguments, and there we use $\rho_{k\ell}(W)$ and $\kappa_{j}(W)$ in place of $\rho_{k\ell}(t)$ and $\kappa_{j}(t)$, where $W$ is the walk (of length $t$) corresponding to $\sigma|_{0:t}$ $\grave{a}$ la Fact \ref{fact:walk}.
%======================================================================

%========================================================================
    \section{Main Results}
\label{s:mainres}
%=========================================================================
    We are now in a position to present our first main result, a proof of which is presented in \S\ref{s:prooft}.
\begin{theorem}
    \label{t:mainres}
        Consider the family of systems \eqref{e:family}. Let $\P_{AS}$ and $\P_{U}\subset\P$ denote the sets of indices of asymptotically stable and unstable systems in \eqref{e:family}, respectively. For each $i,j\in\P$ let the constants $\lambda_{i}$ and $\mu_{ij}$ be as in Fact \ref{fact:key} and \eqref{e:Lyapreln}, respectively. Then the switched system \eqref{e:swsys} is globally asymptotically stable for every switching signal $\sigma$ satisfying
        \begin{align}
            \label{e:thmcondn1} &\li_{t\to+\infty}\:\: \nu(t) > 0 &&\\
            \intertext{and}
            \label{e:thmcondn} &\ls_{t\to+\infty}\:\: \frac{\displaystyle{\sum_{(k,\ell)\in E(\P)}(\ln\mu_{k\ell})\rho_{k\ell}(t) + \sum_{j\in\P_{U}}\abs{\ln\lambda_{j}}\kappa_{j}(t)}}{\displaystyle{\sum_{j\in\P_{AS}}\abs{\ln\lambda_{j}}\kappa_{j}(t)}} < 1,
        \end{align}
        where $\nu(t)$, $\rho_{k\ell}(t)$, and $\kappa_{j}(t)$ are as defined in \eqref{e:swfreq}, \eqref{e:rhokl}, and \eqref{e:kappaj}, respectively.
    \end{theorem}

    \begin{remark}
    \label{r:swfreq}
        The condition \eqref{e:thmcondn1} is necessary to prevent the switched system from eventually ``adhering to" an unstable system. This assumption is natural in our setting because we admit unstable systems in the family \eqref{e:family}, and stipulates that switching continues to occur at a rate that is not asymptotically vanishingly small.
    \end{remark}

    \begin{remark}
    \label{r:thmdescription}
        Condition \eqref{e:thmcondn} involves \emph{only} the asymptotic behaviour of the switching signal. Indeed, \eqref{e:thmcondn} requires that the limit superior of the ratio
        \[
            \displaystyle{\frac{\displaystyle{\sum_{(k,\ell)\in E(\P)}(\ln\mu_{k\ell})\rho_{k\ell}(t) + \sum_{j\in\P_{U}}\abs{\ln\lambda_{j}}\kappa_{j}(t)}}{\displaystyle{\sum_{j\in\P_{AS}}\abs{\ln\lambda_{j}}\kappa_{j}(t)}}}
        \]
        should be strictly less than $1$. In the numerator of the above ratio, the term $\displaystyle{\sum_{(k,\ell)\in E(\P)}(\ln\mu_{k\ell})\rho_{k\ell}(t)}$ captures the number of times each admissible transition $(k,\ell)\in E(\P)$ occurs in $\sigma$ till time $t$, weighted by $\ln\mu_{k\ell}$'s where $\mu_{k\ell}$ is as in Assumption \ref{fact:muij}. The terms $\displaystyle{\sum_{j\in\P_{AS}}\abs{\ln\lambda_{j}}\kappa_{j}(t)}$ and $\displaystyle{\sum_{j\in\P_{U}}\abs{\ln\lambda_{j}}\kappa_{j}(t)}$ capture the number of times a system $j\in\P_{AS}$ (resp. $\P_{U}$) is activated till time $t$ by $\sigma$, weighted by the quantitative measure of (in)stability of the respective system.
    \end{remark}

    \begin{remark}
    \label{r:adt}
        Recall \cite[Theorem 3]{Zhai002} the average dwell time condition for a given family of systems containing unstable subsystems: For any given $\lambda\in\:]\lambda_{1},1[$ there exists a finite constant $\tau_{a}^{*}$ such that the switched system under consideration is globally exponentially stable with stability degree $\lambda$ if the switching signal $\sigma$ satisfies $\displaystyle{\inf_{k>0}\frac{K^{-}(k)}{K^{+}(k)}\ge\frac{\ln\lambda_{2}-\ln\lambda^{*}}{\ln\lambda^{*}-\ln\lambda_{1}}}$ for some scalar $\lambda^{*}\in\:]\lambda_{1},\lambda[$, and the average dwell time is not smaller than $\tau_{a}^{*}$. Here, $K^{-}(k)$ (resp. $K^{+}(k)$) denote the total activation time of Schur stable (resp. unstable) systems; $\lambda_{1}<1$ and $\lambda_{2}\ge 1$ are as follows:
        \begin{align*}
                \norm{A_{i}^{k}} \le
            \begin{cases}
                h_{i}\lambda_{1}^{k},\quad\text{if system $i$ is Schur stable}\\
                h_{i}\lambda_{2}^{k},\quad\text{if system $i$ is unstable}
            \end{cases}
        \end{align*}
        for all $k\ge 1$ with $i = 0,\cdots,N$, and the number of switches \(N_\sigma(0, k)\) on every time interval $0:k-1$, $k\geq 1$, obeys $\displaystyle{N_{\sigma}(0,k) \le N_{0}+\frac{k}{\tau_{a}}}$ with chatter bound $N_{0}$, and average dwell time $\tau_{a}$. Our condition \eqref{e:thmcondn} does not imply nor require any affine bound on the number of switches. Condition \eqref{e:thmcondn} also does not imply nor require a bound on the ratio of the activation time of Schur stable to unstable systems. Consequently, the number of switches $\Ntsigma$ on the interval $1:t$ can grow faster than an affine function of $t$ in our case; indeed, $\Ntsigma$ obeying $k_{0}t-k_{0}'\sqrt{t} \le \Ntsigma \le k_{1}t+k_{1}'t+k_{1}''\sqrt{t}$ for positive constants $k_{0},k_{0}',k_{1},k_{1}',k_{1}''$, is perfectly admissible.
    \end{remark}

    \begin{remark}
    \label{r:uniformity}
        Theorem \ref{t:mainres} does not assert any form of uniformity of global asymptotic stability of \eqref{e:swsys}. Indeed, let $\sigma$ and $\sigma'$ be two switching signals satisfying the hypothesis of Theorem \ref{t:mainres}, and let $(x_{\sigma}(t))_{t\in\N_{0}}$ and $(x_{\sigma'}(t))_{t\in\N_{0}}$ be the corresponding solutions to \eqref{e:swsys}, respectively. Theorem \ref{t:mainres} asserts, in particular, that $x_{\sigma}(t)$ and $x_{\sigma'}(t)\to 0$ as $t\to +\infty$. However, Theorem \ref{t:mainres} does \emph{not} claim that the rates of convergence of the sequences $(x_{\sigma}(t))_{t\in\N_{0}}$ and $(x_{\sigma'}(t))_{t\in\N_{0}}$ are identical.
    \end{remark}

    So far in this section we proposed a class of switching signals corresponding to a given switched system \eqref{e:swsys} such that global asymptotic stability of \eqref{e:swsys} is guaranteed. However, given a switched system \eqref{e:swsys} and a family of numbers $\mu_{ij}$ and $\lambda_{j}$, there may not exist a switching signal $\sigma$ that satisfies condition \eqref{e:thmcondn}. There is also a fair amount of latitude for selection of the Lyapunov functions for the constituent systems in \eqref{e:family}, as is evident from Fact \ref{fact:muij}. Consequently, the sets of numbers $\{\mu_{ij}:(i,j)\in E(\P)\}$ and $\{\lambda_{j}:j\in\P\}$ are not uniquely determined. These numbers enter the condition \eqref{e:thmcondn} in an essential way, and for a certain choice of Lyapunov functions, a.k.a the numbers $\{\mu_{ij}\}$ and $\{\lambda_{j}\}$, it may not be possible to verify\eqref{e:thmcondn} for any switching signal. For instance:
    \begin{example}
    \label{ex:nonex}
        Consider a family of systems with $\P_{AS} = \{1\}$ and $\P_{U} = \{2\}$. Let it be admissible to switch from system $1$ to system $2$ and vice-versa, and let it be not admissible to dwell on any of the systems for two consecutive time steps. Assume $\ln\mu_{12} = -1.5$, $\ln\mu_{21} = 1.8$, $\ln\lambda_{1} = -0.2$, and $\ln\lambda_{2} = 1.6$. In this case a switch occurs at every time $t$. The term
        \begin{align*}
            (\ln\mu_{12})\rho_{12}(t) + (\ln\mu_{21})\rho_{21}(t) + \abs{\ln\lambda_{2}}\kappa_{2}(t)\\
            = -1.5\rho_{12}(t) + 1.8\rho_{21}(t) + 1.6\kappa_{2}(t)
        \end{align*}
        Consequently, condition \eqref{e:thmcondn} is not satisfied.
    \end{example}
    In view of Example \ref{ex:nonex}, given a switched system \eqref{e:swsys}, an important and natural problem concerns the existence of a switching signal $\sigma$ that satisfies condition \eqref{e:thmcondn}. We address this problem in the remainder of this section.

    For a streamlined presentation of our second main result, we employ from \S\ref{ss:swprop} the directed graph $G(\P,E(\P))$ representation of the switched system \eqref{e:swsys}. Given the family of systems \eqref{e:family}, we get an estimate of the constants $\mu_{k\ell}$ for all $(k,\ell)\in E(\P)$ and $\lambda_{j}$ for all $j\in\P$, by applying Proposition \ref{p:muprop} and Fact \ref{fact:key}, respectively. The problem at hand, can now be rephrased in a purely graph theoretic language, as:
     \begin{problem}
     \label{prob:graphprob}
        Given a directed graph $G(\P,E(\P))$ and two sets of real numbers $\{\mu_{k\ell}:(k,\ell)\in E(\P)\}$, and $\{\lambda_{j}:j\in\P\}$, does there exist an infinite walk on $G(\P,E(\P))$ such that the corresponding switching signal $\grave{a}$ la Fact \ref{fact:walk} satisfies condition $\eqref{e:thmcondn}$?
     \end{problem}

    \begin{remark}
     \label{r:trivcase}
        On the one hand, the issue of existence of a switching signal that satisfies condition \eqref{e:thmcondn} is trivial if the directed graph $G(\P,E(\P))$ has a self-loop for any vertex $j\in\P_{AS}$. Indeed, for a switching signal whose corresponding infinite walk ($\grave{a}$ la Fact \ref{fact:walk}) traverses this self-loop repeatedly, the denominator of the ratio on the left-hand side of \eqref{e:thmcondn} tends to $+\infty$ as $t$ (a.k.a, the length of the walk) tends to $+\infty$. In the numerator of \eqref{e:thmcondn}, $\ln\mu_{jj} = \ln(\lambda_{\max}(P_{j}P_{j}^{-1})) = 0$ by Proposition \ref{p:muprop}, and consequently, condition \eqref{e:thmcondn} holds. The design of an algorithm to detect such a walk is also simple: given $G(\P,E(\P))$, the algorithm needs to detect the vertex corresponding to an asymptotically stable subsystem with a self-loop. Beyond the preceding trivial case, on the other hand, given a weighted directed graph, the problem of algorithmically finding an \emph{infinite} walk that satisfies some pre-specified conditions involving vertex and edge weights in the form of the numbers $\mu_{ij}$ and $\lambda_{j}$ in Problem \ref{prob:graphprob}, is not a straightforward task.
     \end{remark}

      Against the backdrop of Remark \ref{r:trivcase}, we propose a partial solution to Problem \ref{prob:graphprob} in the sequel: we provide a sufficient condition for the existence of a switching signal $\sigma$ that satisfies condition \eqref{e:thmcondn}, and an algorithm to design such a switching signal.

     Recall \cite[p.6]{Harris} that a walk on a graph is called a \emph{trail} if all its edges are distinct. A closed trail is called a \emph{circuit}.

     Let $A = [a_{ij}]$ be the (node arc) incidence matrix \cite[\S3.4]{papa_optimization} of $G(\P,E(\P))$, defined by
    \begin{align}
    \label{e:incimat}
        a_{ij} = \begin{cases}
                +1, & \text{if edge $(i,j)$ leaves node $i$, $i = 1,2,\cdots,\abs{\P}$},\\
                -1, & \text{if edge $(i,j)$ enters node $i$, $j = 1,2,\cdots,\abs{E(\P)}$},\\
                0, & \text{otherwise},
                \end{cases}
    \end{align}
    where $\abs{S}$ denotes the cardinality of a finite set $S$.

    \begin{remark}
    \label{r:self-loop}
        Incidence matrices are commonly defined for graphs without self-loops. Our purposes require us to define incidence matrices for directed graphs with self-loops, and we accommodate the latter in an incidence matrix $A$ in the following way: Suppose that the given directed graph $G(\P,E(\P))$ has a self-loop on vertex $j\in\P$. We consider an auxiliary vertex $j'$ corresponding to $j$ and represent the self-loop as an edge from $j$ to $j'$. Consequently, the number of rows of $A$ becomes $\abs{\P}+\abs{\text{the set of vertices having self-loops}}$. Here is an illustration of this procedure:
    \end{remark}

    \begin{example}
    \label{ex:inciex}
        Consider the directed graph below:
    \begin{center}
        \begin{tikzpicture}[every path/.style={>=latex},every node/.style={auto}]
            \node[state]            (a) at (-1.5,0)  { 1 };
            \node[state]            (b) at (1.5,0)  { 2 };
            \node[state]            (c) at (0,-1.5) { 3 };

            \draw[->] (a) edge[loop left] (a);
            \draw[->] (a) edge (b);
            \draw[->] (b) edge[bend right] (a);
            \draw[->] (a) edge (c);
            \draw[->] (b) edge (c);
        \end{tikzpicture}
    \end{center}
    for which the incidence matrix is:
    \begin{align*}
        A =
        \bordermatrix{&\text{$(1,1')$} &\text{$(2,1)$} & \text{$(1,2)$} & \text{$(2,3)$} & \text{$(1,3)$} \cr
                      \text{$1$}&+1 & -1 & +1 & 0 & +1 \cr
                      \text{$1'$}&-1 & 0 & 0 & 0 & 0\cr
                      \text{$2$}&0 & +1 & -1 & +1 & 0\cr
                      \text{$3$}&0 & 0 & 0 & -1 & -1}.
    \end{align*}
    \end{example}

    Our second main result is as follows:
    \begin{theorem}
    \label{t:graphres}
        Consider the switched system \eqref{e:family}, and the directed graph $G(\P,E(\P))$ as in \S\ref{ss:swprop}. Let the sets of real numbers $\{\mu_{k\ell}:(k,\ell)\in E(\P)\}$ and  $\{\lambda_{j}:j\in\P\}$ be calculated from Fact \ref{fact:key} and Proposition \ref{p:muprop}, respectively.
        \renewcommand{\theenumii}{\alph{enumii}}
        \begin{enumerate}[label = (\alph*)]
        \item \label{t2:stepa} If there exists a closed walk $W$ on $G(\P,E(\P))$ that satisfies the inequality:
        \begin{align}
        \label{e:graphcondn}
            \frac{\displaystyle{\sum_{(k,\ell)\in E(\P)}(\ln\mu_{k\ell})\rho_{k\ell}(W) + \sum_{j\in\P_{U}}\abs{\ln\lambda_{j}}\kappa_{j}(W)}}{\displaystyle{\sum_{j\in\P_{AS}}\abs{\ln\lambda_{j}}\kappa_{j}(W)}} < 1,
        \end{align}
         then the switching signal corresponding to the infinite walk --- obtained by repeating the closed walk $W$ --- satisfies \eqref{e:thmcondn}.\footnote{The correspondence refers to Fact \ref{fact:walk}.}
        \item \label{t2:stepb} The existence of a circuit on $G(\P,E(\P))$ satisfying \eqref{e:graphcondn} is guaranteed by the existence of a solution to the following feasibility problem (linear program) in the variable $f\in\R^{\abs{E(\P)}}$:\\
        \begin{align}
        \label{e:optprob}
            \minimize &\quad 1 \\
            \sbjto &\quad
            \begin{cases}
                Af = (0,0,\cdots,0)^\top,& \nn\\
                \text{condition \eqref{e:graphcondn}}, &\nn\\
                0 \le f_{j} \le 1\:\:\text{for all}\:\: 1\le j \le \abs{E(\P)}, &\nn\\
                \displaystyle{\sum_{j=1}^{\abs{E(\P)}}f_{j}} \geq 1,& \nn
            \end{cases}
        \end{align}
        where the matrix $A$ is as defined in \eqref{e:incimat}, and $\abs{E(P)}$ denotes the cardinality of the set $E(\P)$.
        \item \label{t2:stepc} If a solution to the feasibility problem \eqref{e:optprob} exists, then Hierholzer's algorithm can be used to obtain the circuit on $G(\P,E(\P))$ that satisfies \eqref{e:graphcondn}.%\footnote{See \cite[p.57]{Harris} for Hierholzer's algorithm.}
    \end{enumerate}
    \end{theorem}

    A detailed proof of Theorem \ref{t:graphres} is provided in \S\ref{s:prooft}.

    \begin{remark}
    \label{r:thm2description}
        Theorem \ref{t:graphres} gives a \emph{sufficient condition} for the existence of an infinite walk on $G(\P,E(\P))$ corresponding to a switching signal $\sigma$ that satisfies condition \eqref{e:thmcondn}, thereby providing an answer to Problem \ref{prob:graphprob}. Given a directed graph representing the switched system \eqref{e:swsys}, finding a \emph{necessary condition} for the directed graph to admit an infinite walk such that condition \eqref{e:thmcondn} holds for the corresponding switching signal is a difficult problem. Armed with the sufficient condition proposed in Theorem \ref{t:graphres}, we need to algorithmically determine a closed walk on $G(\P,E(\P))$ that satisfies \eqref{e:graphcondn}. We solve the last problem in two steps: Firstly, we employ the feasibility problem \eqref{e:optprob} to determine whether there exists a circuit on $G(\P,E(\P))$ that satisfies \eqref{e:graphcondn} by algorithmically calculating the vector $f$, each entry of which is either one or zero, corresponding to whether an edge is included in the circuit or not, respectively. If the feasibility problem \eqref{e:optprob} has a solution, we proceed to the second step with the calculated vector $f$, and apply Hierholzer's algorithm to find the circuit on $G(\P,E(\P))$ that satisfies \eqref{e:graphcondn}. See also Remark \ref{r:algo} below.
    \end{remark}

    \begin{remark}
         We mentioned in Remark \ref{r:thm2description} that in a solution $f$ to the feasibility problem \eqref{e:optprob}, (if there is a solution,) each entry of the column vector $f$ is either one or zero. This implies that the length of a circuit obtained as a solution to the feasibility problem \eqref{e:optprob} can at most be the total number of edges of the directed graph $G(\P,E(\P))$, i.e., the case when all entries of the vector $f$ are one.
    \end{remark}

    \begin{remark}
    \label{r:connection}
        Condition \eqref{e:graphcondn} is included in the feasibility problem \eqref{e:optprob} in the following manner: We assume that the total number of times the closed walk $W$ visits a vertex $j\in\P$ is the same as the total number of times $W$ visits the outgoing edges of the vertex $j$. Consequently, for a vertex $j\in\P$, $\kappa_{j}(W)$ can be replaced by $\rho_{j\ell}(W)$, $(j,\ell)\in E(\P)$. Since we are concerned with an infinite walk constructed by repeating the closed walk $W$ satisfying \eqref{e:graphcondn}, this assumption is no loss of generality.
    \end{remark}

    \begin{remark}
    \label{r:zerovec}
        The condition $\displaystyle{Af = (0,0,\cdots,0)^\top}$ in the feasibility problem \eqref{e:optprob} in Theorem \ref{t:graphres} represents a closed walk. As such, the preceding equality always has a trivial solution where $f$ is a vector with all entries equal to $0$. The condition $\displaystyle{\sum_{j}f_{j} \geq 1}$ in \eqref{e:optprob} ensures that any solution $f$ to \eqref{e:optprob} is not the zero vector.
    \end{remark}

    \begin{remark}
    \label{r:algo}
        Given an Eulerian graph $G$, Hierholzer's algorithm \cite[p.57]{Harris} finds an Eulerian circuit of $G$; see (c) of Theorem \ref{t:graphres}. The applicability of this algorithm in our context is explained in the proof of Theorem \ref{t:graphres} in \S\ref{s:prooft}.
    \end{remark}
%==========================================================================

%===========================================================================
\section{Numerical Example}
\label{s:ex}
%=============================================================================
    Consider a family of systems \eqref{e:family} with $\P = \{1,2,3,4,5\}$, and
    \begin{align*}
        A_{1} &= \pmat{0.4 & 0.8\\-0.7 & 0.6},& A_{2} &= \pmat{0.3 & 0.6\\0.1 & 0.4},\\
        A_{3} &= \pmat{1 & 0\\0 & 0.5},& A_{4} &= \pmat{1.2 & 0.7\\1.6 & 0.1},\\
        A_{5} &= \pmat{1 & 0.1\\0.1 & 1}.
    \end{align*}
     For this family $\P_{AS}=\{1,2\}$ and $\P_{U}=\{4,5\}$. Let all transitions among the systems in the given family be admissible. Let it also be permissible for switching signals to dwell on systems $3$, $4$, and $5$ for at least two consecutive time steps.
    \begin{center}
    \begin{tikzpicture}[every path/.style={>=latex},every node/.style={auto}]
            \node[state] (a) at (-2,2)   {$1$};
            \node[state] (b) at (2,2)  {$2$};
            \node[state] (c) at (-1,0) {$3$};
            \node[state] (d) at (1,0) {$4$};
            \node[state] (e) at (0,-2) {$5$};

            \draw[->] (a) edge (b);
            \draw[->] (a) edge (c);
            \draw[->] (a) edge (d);
            \draw[->] (a) edge [bend right = 60] (e);

            \draw[->] (b) edge [bend right] (a);
            \draw[->] (b) edge (c);
            \draw[->] (b) edge (d);
            \draw[->] (b) edge [bend left = 60] (e);

            \draw[->] (c) edge [bend left] (a);
            \draw[->] (c) edge [bend left] (b);
            \draw[->] (c) edge (d);
            \draw[->] (c) edge (e);

            \draw[->] (d) edge [bend right] (a);
            \draw[->] (d) edge [bend right] (b);
            \draw[->] (d) edge [bend left](c);
            \draw[->] (d) edge (e);

            \draw[->] (e) edge [bend left = 45] (a);
            \draw[->] (e) edge [bend right = 45] (b);
            \draw[->] (e) edge [bend left](c);
            \draw[->] (e) edge [bend right] (d);

            \draw[->] (c) edge [loop left] (c);
            \draw[->] (d) edge [loop right] (d);
            \draw[->] (e) edge [loop below] (e);
        \end{tikzpicture}
        \end{center}
    That is,
    \begin{align*}
        \P &= \{1,2,3,4,5\}, \nn\\
        \intertext{and}
        E(\P) &= \{(1,2),(1,3),(1,4),(1,5),\\
              &\qquad(2,1),(2,3),(2,4),(2,5),\\
              &\qquad(3,1),(3,2),(3,3'),(3,4),(3,5),\\
              &\qquad(4,1),(4,2),(4,3),(4,4'),(4,5),\\
              &\qquad(5,1),(5,2'),(5,3),(5,4),(5,5')\},
    \end{align*}
    where $3',4'$ and $5'$ are as explained in Remark \ref{r:self-loop}. We design the incidence matrix $A$ from the above description of $E(\P)$ and associate the elements of the column vector $f$ with the entries of $E(\P)$ as illustrated in Example \ref{ex:inciex}. (We do not provide the rather large incidence matrix here for reasons of space.)

    For the given family of systems \eqref{e:family}, an estimate of the pairs $(P_{i},\lambda_{i})$ as in Fact \ref{fact:key} are as follows:
    \begin{align*}
        (P_{1},\lambda_{1}) &= \Biggl(\pmat{4.7545 & -0.5804\\-0.5804 & 5.4464},0.8269\Biggr),&&\\
        (P_{2},\lambda_{2}) &= \Biggl(\pmat{1.1421 & 0.3422\\0.3422 & 1.8755},0.5026\Biggr),&&\\
        (P_{3},\lambda_{3}) &= \Biggl(\pmat{1 & 0\\0 & 1}, 1\Biggr),&&\\
        (P_{4},\lambda_{4}) &= \Biggl(\pmat{1 & 0\\0 & 1}, 5.1306\Biggr),&&\\
        (P_{5},\lambda_{5}) &= \Biggl(\pmat{1 & 0\\0 & 1}, 3.2000\Biggr).
    \end{align*}
    From Proposition \ref{p:muprop}, we obtain the following estimates for $\mu_{ij}$:
    \begin{align*}
        \mu_{12} &= 0.4185 , & \mu_{13} &= 0.2260 ,& \mu_{14} &= 0.2260 , &\\
        \mu_{15} &= 0.2260 , & \mu_{21} &= 5.2823 ,& \mu_{23} &= 0.9928 , &\\
        \mu_{24} &= 0.9928 , & \mu_{25} &= 0.9928 , & \mu_{31} &= 5.7761 , &\\
        \mu_{32} &= 2.0103 , & \mu_{33} &= 1 , & \mu_{34} &= 1 , &\\
        \mu_{35} &= 1 ,& \mu_{41} &= 5.7761 , & \mu_{42} &= 2.0103 , &\\
        \mu_{43} &= 1 ,&\mu_{44} &= 1 , & \mu_{45} &= 1 , &\\
        \mu_{51} &= 5.7761 ,&\mu_{52} &= 2.0103 , & \mu_{53} &= 1 , &\\
        \mu_{54} &= 1 ,& \mu_{55} &= 1.
    \end{align*}

    Solving the feasibility problem \eqref{e:optprob} in the context of this example with the aid of MATLAB by employing the program YALMIP \cite{Lofberg04} and the solver SDPT3, we obtain the following solution:
    \begin{align*}
        f &= (1,1,0,0,1,1,0,0,1,1,0,0,0,0,0,0,0,0,0,0,0,0,0)^\top,
    \end{align*}
    with
    \[
        \frac{\displaystyle{\sum_{(k,\ell)\in E(\P)}(\ln\mu_{k\ell})\rho_{k\ell}(W) + \sum_{j\in\P_{U}}\abs{\ln\lambda_{j}}\kappa_{j}(W)}}{\displaystyle{\sum_{j\in\P_{AS}}\abs{\ln\lambda_{j}}\kappa_{j}(W)}} = 0.99 < 1.
    \]
    Following is a circuit obtained from the vector $f$ with the aid of Hierholzer's algorithm:
    \begin{align*}
        3,(3,1),1,(1,2),2,(2,1),1,(1,3),3,(3,2),2,(2,3),3,
    \end{align*}
    which pictorially is as follows:
        \begin{center}
        \usetikzlibrary{automata,positioning}
        \begin{tikzpicture}[every path/.style={>=latex},every node/.style={auto}]
            \node[state] (1) {$1$};
            \node[state] (2) [right=of 1] {$2$};
            \node[state] (3) [right=of 2] {$3$};

            \path[->] (3) edge [bend left = 50] node {$a$} (1)
                      (1) edge node {$b$} (2)
                      (2) edge [bend left] node {$c$} (1)
                      (1) edge [bend left] node {$d$} (3)
                      (3) edge [bend left] node {$e$} (2)
                      (2) edge  node {$f$} (3);
        \end{tikzpicture}
        \end{center}
        Here $a,b,c,d$ and $e$ denote the edges in consecutive order in which they appear in the above circuit.

\begin{figure}[htbp]
    \begin{center}
        \includegraphics[scale=0.5]{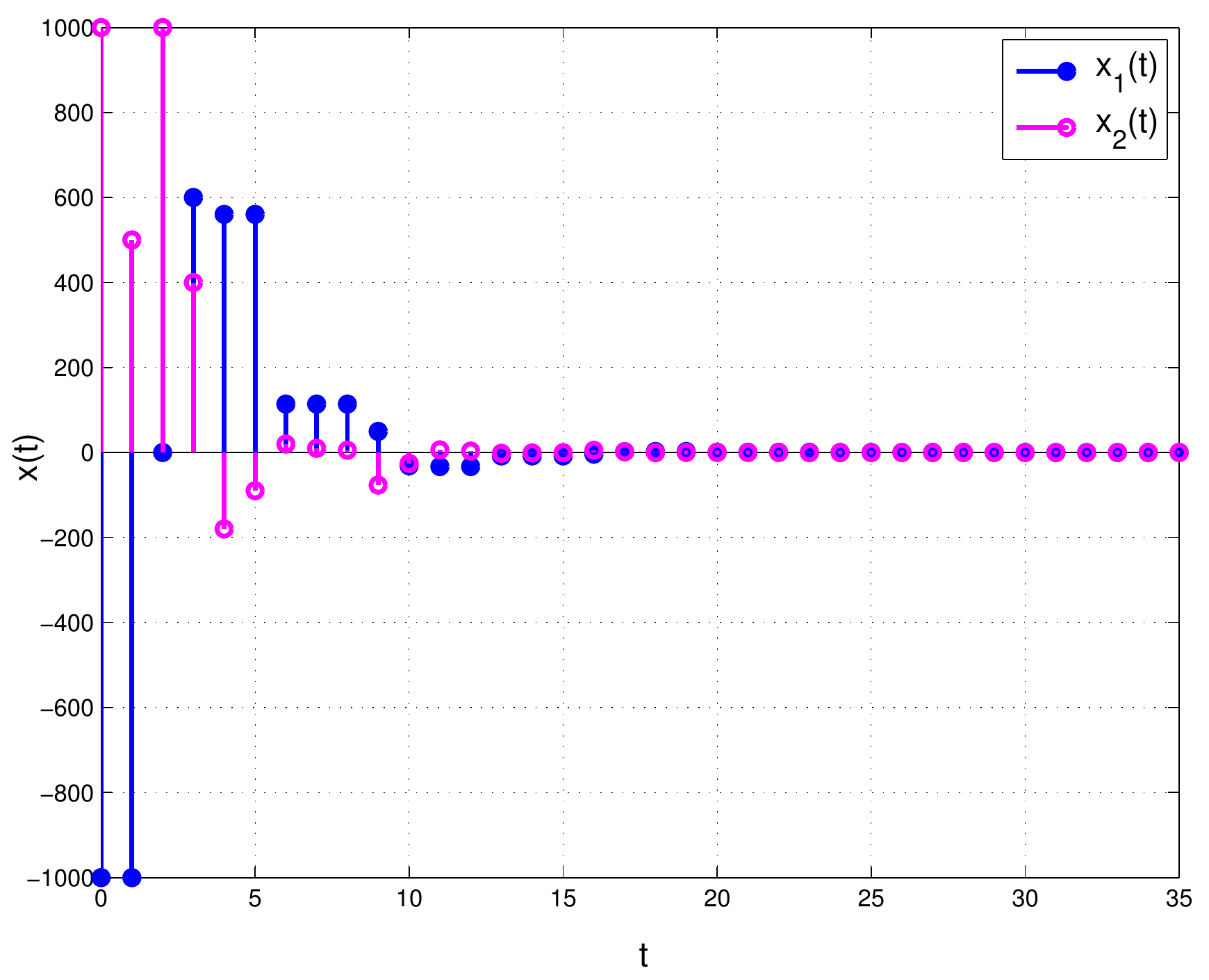}
        \caption{Solution to \eqref{e:swsys} with $x_{1}(0) = -1000, x_{2}(0) = 1000$.}
        \label{f:response1}
    \vspace*{0.5cm}
        \includegraphics[scale=0.5]{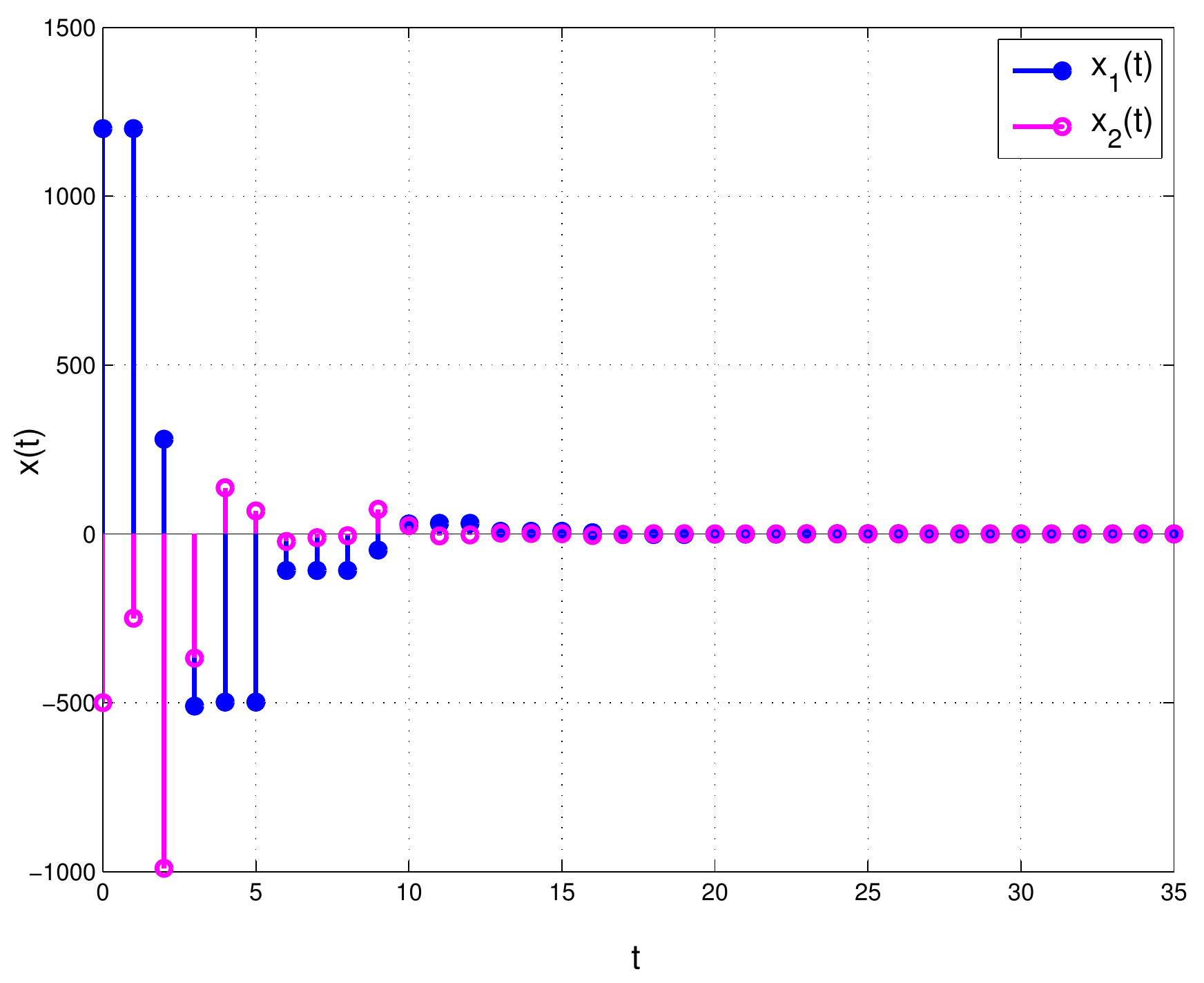}
        \caption{Solution to \eqref{e:swsys} with $x_{1}(0) = 1200, x_{2}(0) = -500$.}
        \label{f:response2}
    \end{center}
    \end{figure}

    We study the nature of $(x(t))_{t\in\N_{0}}$ for different initial conditions $x_{0}$ and observe that in each case, $x(t) \to 0$ as $t \to +\infty$. In Figures \ref{f:response1} and \ref{f:response2} we present two representative solutions to \eqref{e:swsys} with $x_{0} = (-1000,1000)^\top$ and $x_{0} = (1200,-500)^\top$, respectively.

%=================================================================================

%====================================================================================
\section{Conclusion}
\label{s:concln}
%====================================================================================
    In this article we proposed a class of switching signals under which global asymptotic stability of a discrete-time switched linear system is guaranteed. We considered unstable dynamics in the family and our characterization of the stabilizing switching signal involved only asymptotic behaviour of the switching signal. Moreover, given a switched system, we proposed a sufficient condition for the existence of a switching signal that satisfies our condition. We presented an algorithm to design such a switching signal.

    Several open questions in this area remain. A necessary condition for the existence of the proposed switching signal is currently under investigation and will be reported elsewhere. A randomized algorithm for the synthesis of stabilizing switching signals for large-scale switched systems is also being developed.
%==========================================================================================

%==========================================================================================
\section{Proofs}
\label{s:allproofs}
%===========================================================================================
%==========================================================================
\subsection{Proofs of Fact \ref{fact:key}, Proposition \ref{p:muprop}, and Fact \ref{fact:walk}}
\label{s:proofs}
%============================================================================
\ \\
    \begin{proof}[of Fact \ref{fact:key}]
        For asymptotically stable systems let $\R^{d}\ni z\longmapsto V_{i}(z) \Let z^\top P_{i}z$, where $P_{i}\in\R^{d\times d}$ is the symmetric and positive definite solution to the discrete-time Lyapunov equation
        \begin{align}
        \label{e:dtlyapeqn}
            A_{i}^\top P_{i}A_{i} - P_{i} + Q_{i} = 0
        \end{align}
        for some pre-selected symmetric and positive definite matrix $Q_{i}\in\R^{d\times d}$ \cite[Proposition 11.10.5]{Bernstein}. If $A_{i}$ is marginally stable, it is known \cite[Proposition 11.10.6]{Bernstein} that there exists a symmetric and positive definite matrix $P_{i}\in\R^{d\times d}$ and a symmetric and non-negative definite matrix $Q_{i}\in\R^{d\times d}$ that solve the discrete-time Lyapunov equation \eqref{e:dtlyapeqn}; we put $\R^{d}\ni z\longmapsto V_{i}(z) \Let z^\top P_{i}z$ as the corresponding Lyapunov-like function. A straightforward calculation gives
        \[
            V_{i}(z(t+1)) - V_{i}(z(t)) = -z(t)^\top Q_{i}z(t)
        \]
        in both the cases. An application of the standard inequality \cite[Lemma 8.4.3]{Bernstein} leads to
        \[
            -z^\top Q_{i}z \le -\frac{\lambda_{\min}(Q_{i})}{\lambda_{\max}(P_{i})}z^\top P_{i}z.
        \]
        Defining $\bar{\lambda}_{i} = \frac{\lambda_{\min}(Q_{i})}{\lambda_{\max}(P_{i})}$, we arrive at
        \[
            V_{i}(z(t+1)) \le \lambda_{i}V_{i}(z(t)) \:\: \text{with} \:\: \lambda_{i} = 1 - \bar{\lambda}_{i},
        \]
        which gives \eqref{e:Lyapprop} with $0 < \lambda_{i} \le 1$.\\
        For unstable systems, let us consider the simplest case of a symmetric and positive definite matrix $P_{i} = I_{d\times d}$, and let $\R^{d}\ni z\longmapsto V_{i}(z) \Let \norm{z}^{2}$. Then by the Cauchy-Schwarz inequality and sub-multiplicativity property of matrix norms,
        \[
            \abs{V_{i}(z(t+1))-V_{i}(z(t))} \leq 2\norm{A_{i}}V_{i}(z(t)).
        \]
        To wit,
        \[
            -2\norm{A_{i}}V_{i}(z(t)) \leq V_{i}(z(t+1))-V_{i}(z(t)) \leq  2\norm{A_{i}}V_{i}(z(t))
        \]
        for all $t\in\N_{0}$.
        Thus \eqref{e:Lyapprop} holds for unstable systems with $\lambda_{i} = 1+2\norm{A_{i}} > 1$.
    \end{proof}

    \begin{proof}[of Proposition \ref{p:muprop}]
        Observe that $P_{j}{P_{i}}^{-1}$ is similar to $P_{i}^{-1/2}(P_{j}P_{i}^{-1})P_{i}^{1/2}$, and that the matrix $P_{i}^{-1/2}P_{j}P_{i}^{-1/2}$ is symmetric and positive definite. Since the spectrum of a matrix is invariant under similarity transformations, the eigenvalues of $P_{j}{P_{i}}^{-1}$ are the same as the eigenvalues of $P_{i}^{-1/2}P_{j}P_{i}^{-1/2}$; consequently, the eigenvalues of $P_{j}{P_{i}}^{-1}$ are real numbers. In addition,
        \[
            \displaystyle{\sup_{0\neq z\in\R^{d}}\frac{\langle P_{j}z,z\rangle}{\langle P_{i}z,z\rangle} = \sup_{0\neq z\in\R^{d}}\frac{\langle P_{j}z,z\rangle}{\langle P_{i}^{1/2}z,P_{i}^{1/2}z\rangle}},
        \]
    and, with $z \Let P_{i}^{-1/2}y$, the right-hand side above is
    \begin{align*}
        \sup_{0\neq y\in\R^{d}}\frac{\langle P_{j}(P_{i}^{-1/2}y),P_{i}^{-1/2}y\rangle}{\langle y,y\rangle}
       	& = \sup_{0\neq y\in\R^{d}}\frac{\langle P_{i}^{-1/2}P_{j}P_{i}^{-1/2}y,y\rangle}{\langle y,y\rangle}\\
       	& = \lambda_{\max}(P_{i}^{-1/2}P_{j}P_{i}^{-1/2})\\
       	&= \lambda_{\max}(P_{j}P_{i}^{-1}).
    \end{align*}
    	Since ${V_{j}(z)}\leq\mu_{ij}{V_{i}(z)}$ for all $z\in\R^{d}$, the smallest constant $\mu_{ij}$ satisfies \eqref{e:muij}.
    \end{proof}

    \begin{proof}[of Fact \ref{fact:walk}]
        Let ${\P}^{\N_{0}}$ denote the set of all sequences $(x_{n})_{n\in\N_{0}}$ with $x_{n}\in\P$. Define the map
        \[
            \Phi:\:{\P}^{\N_{0}}\to\{W:\text{$W$ is an infinite walk on $G(\P,E(\P))$}\}
        \]
        by $\sigma \longmapsto \Phi(\sigma) \Let \bigl(\sigma(0), (\sigma(0),\sigma(1)), \sigma(1), (\sigma(1),\sigma(2)), \sigma(2)$, $\cdots\bigr)$, which is a unique infinite walk on $G(\P,E(\P))$. Indeed, for two different switching signals $\sigma_{1}$ and $\sigma_{2}$ the terms $\sigma_{1}(\ell),(\sigma_{1}(\ell),\sigma_{1}(\ell+1)),\sigma_{1}(\ell+1)$ and $\sigma_{2}(\ell),(\sigma_{2}(\ell),\sigma_{2}(\ell+1)),\sigma_{2}(\ell+1)$ are different at least for one $\ell\in\N_{0}$. Consequently, $\Phi(\sigma_{1})$ and $\Phi(\sigma_{2})$ are different, which ensures injectivity of $\Phi$. Surjectivity is clear.\par
        Conversely, for an infinite walk $W = \bigl(a_{0},(a_{0},a_{1}),a_{1},(a_{1}, a_{2})$, $\cdots\bigr)$, we define $\sigma:\:\N_{0}\to\P$ by $\sigma(i)=a_{i}$, $(a_{i},a_{i+1})$ be the $i$-th transition of $\sigma$, i.e., $\sigma(i) = a_{i}, \sigma(i+1)=a_{i+1}, i = 0,1,2,\cdots$. Define the map
        \[
            \Psi:\:\{W:\text{$W$ is an infinite walk on $G(\P,E(\P))$}\} \to \P^{\N_{0}}
        \]
        by $W \longmapsto \Psi(W) \Let \sigma$, which is the switching signal corresponding to the infinite walk $W$ constructed above. Standard arguments as above may be employed to assert bijectivity of $\Psi$. Finally, it is clear that $\Phi$ and $\Psi$ are inverses of each other.
    \end{proof}

%====================================================================
\subsection{Proofs of Theorems \ref{t:mainres} and \ref{t:graphres}}
\label{s:prooft}
%=====================================================================
\ \\
\begin{proof}[of theorem \ref{t:mainres}]
    We will employ properties of the Lyapunov-like functions $V_{i}$ for all $i\in\P$. Fix $t>0$. Recall from \S\ref{s:prelim} that $0\teL\tau_{0}<\tau_{1}<\cdots<\tau_{\Ntsigma}$ are the switching instants before (and including) $t$. A straightforward iteration involving Fact \ref{fact:key} and Fact \ref{fact:muij} leads to
    \begin{align}
    \label{e:proof1}
        V_{\sigma(t)}(x(t)) &\leq V_{\sigma(0)}(x_{0})\times\nn\\
                            &\biggl({\prod_{i=0}^{\Ntsigma-1}\mu_{\sigma(\tau_{i})\sigma(\tau_{i+1})}}{\prod_{i=0}^{\Ntsigma-1}\lambda_{\sigma(\tau_{i})}^{S_{i+1}}}{\lambda_{\sigma(\Ntsigma)}^{t-\tau_{\Ntsigma}}}\biggr),
    \end{align}
    where $S_{i+1}$ is as defined in \eqref{e:holdtime}. The second term on the right-hand side of the above inequality can be written as
    \begin{align*}
        \exp\biggl(\ln{\biggl(\prod_{i=0}^{\Ntsigma-1}\mu_{\sigma(\tau_{i})\sigma(\tau_{i+1})}\biggr)}+\ln{\biggl(\prod_{i=0}^{\Ntsigma-1}\lambda_{\sigma(\tau_{i})}^{S_{i+1}}\biggr)}&\\
        +\ln{\lambda_{\sigma(\tau_{\Ntsigma})}^{t-\tau_{\Ntsigma}}}\biggr).&
    \end{align*}
    Now
    \begin{align}
    \label{e:proof2}
        \ln{\biggl(\prod_{i=0}^{\Ntsigma-1}\mu_{\sigma(\tau_{i})\sigma(\tau_{i+1})}\biggr)} &= \sum_{i=0}^{\Ntsigma-1}\ln\mu_{\sigma(\tau_{i})\sigma(\tau_{i+1})} \nn\\
        &= \sum_{k\in\P}\sum_{i=0}^{\Ntsigma-1}\sum_{\substack{k\to\ell:\\\ell\in\P,\\k\neq\ell,\\\sigma(\tau_{i})=k,\\\sigma(\tau_{i+1})=\ell}}\ln\mu_{k\ell} \nn\\
        &= \sum_{(k,\ell)\in E(\P)}(\ln\mu_{k\ell})\rho_{k\ell}(t),
    \end{align}
    where $\rho_{k\ell}(t)$ is as defined in \eqref{e:rhokl}, and
    \begin{align*}
        \ln{\biggl(\prod_{i=0}^{\Ntsigma-1}\lambda_{\sigma(\tau_{i})}^{S_{i+1}}\biggr)} &= \sum_{i=0}^{\Ntsigma-1}S_{i+1}\ln\lambda_{\sigma(\tau_{i})}\nn\\
        &= \sum_{i=0}^{\Ntsigma-1}\biggl(\sum_{j\in\P}\indicator S_{i+1}\ln\lambda_{j}\biggr).
    \end{align*}
    Separating out the asymptotically stable, marginally stable, and unstable subsystems in the family \eqref{e:family} into the subsets $\P_{AS}$, $\P_{MS}$, and $\P_{U}\subset\P$, respectively, we see that the right-hand side of the above equation can be written as
    \begin{align*}
        \sum_{j\in\P_{AS}}\ln\lambda_{j}\sum_{\substack{i:\sigma(\tau_{i})=j\\i=0,1,\cdots,\Ntsigma}}S_{i+1} &+ \sum_{j\in\P_{MS}}\ln\lambda_{j}\sum_{\substack{i:\sigma(\tau_{i})=j\\i=0,1,\cdots,\Ntsigma}}S_{i+1}\\
        &+ \sum_{j\in\P_{U}}\ln\lambda_{j}\sum_{\substack{i:\sigma(\tau_{i})=j\\i=0,1,\cdots,\Ntsigma}}S_{i+1}.
    \end{align*}
    Recall from Fact \ref{fact:key} that $0<\lambda_{j}<1$, $\lambda_{j}=1$, and $\lambda_{j}>1$ for the first, middle, and the last sums, respectively. Thus the last expression above equals
    \begin{align}
    \label{e:proof3}
        -\sum_{j\in\P_{AS}}\abs{\ln\lambda_{j}}\kappa_{j}(t) + \sum_{j\in\P_{U}}\abs{\ln\lambda_{j}}\kappa_{j}(t),
    \end{align}
    where $\kappa_{j}(t)$ is as defined in \eqref{e:kappaj}.
    We define two functions
    \begin{align}
    \label{e:g1defn}
        \N\ni t\longmapsto g_{1}(t) \Let \sum_{j\in\P_{AS}}\abs{\ln\lambda_{j}}\kappa_{j}(t),
    \end{align}
    and
    \begin{align}
    \label{e:g2defn}
        \N\ni t\longmapsto g_{2}(t) \Let \sum_{(k,\ell)\in E(\P)}(\ln\mu_{k\ell})\rho_{k\ell}(t) &\\+ \sum_{j\in\P_{U}}\abs{\ln\lambda_{j}}\kappa_{j}(t) + (t-\tau_{\Ntsigma})\ln\lambda_{\sigma(\tau_{\Ntsigma})}.
    \end{align}
    Substituting \eqref{e:proof2} and \eqref{e:proof3} in \eqref{e:proof1} and applying the definitions of $g_{1}$ and $g_{2}$ from \eqref{e:g1defn} and \eqref{e:g2defn}, respectively, we obtain
    \begin{align}
    \label{e:proof4}
        V_{\sigma(t)}(x(t)) \le V_{\sigma(0)}(x_{0})\exp{\bigl(g_{2}(t)-g_{1}(t)\bigr)}.
    \end{align}

    To verify GAS of the switched system \eqref{e:swsys}, (by Definition \ref{d:gas}), we need to find conditions such that
    \begin{align}
        &\lim_{t\to+\infty}\exp{\bigl(g_{2}(t)-g_{1}(t)\bigr)} = 0, \label{e:gascondn1} \\
        \intertext{and that}
        &\text{convergence is uniform for initial conditions $\tilde{x}_{0}$ satisfying} \nn\\
        &\norm{\tilde{x}_{0}} \le \norm{x_{0}}. \label{e:gascondn2}
    \end{align}
    A sufficient condition for \eqref{e:gascondn1} is that
    \begin{align}
    \label{e:condn1suff}
        \li_{t\to +\infty}\bigl(g_{1}(t)-g_{2}(t)\bigr) = +\infty,
    \end{align}
    so our proof will be complete if we establish \eqref{e:condn1suff} and verify \eqref{e:gascondn2} separately. The following steps are geared towards establishing \eqref{e:condn1suff}.

    The hypothesis \eqref{e:thmcondn1} guarantees that $t-\tau_{\Ntsigma}$ in the definition of $g_{2}$ in \eqref{e:g2defn} is $o(t)$ as $t\to+\infty$. Indeed, $t-\tau_{\Ntsigma}\ne o(t)$ as $t\to +\infty$ implies that $\li_{t\to +\infty}\nu(t) = 0$, which contradicts our assumption \eqref{e:thmcondn1}. Thus, we can omit this term in further analysis. Next, condition \eqref{e:condn1suff} is equivalent to
    \begin{align}
    \label{e:condn1suff1}
        \li_{t\to+\infty}\Biggl(g_{1}(t)\biggl(1-\frac{g_{2}(t)}{g_{1}(t)}\biggr)\Biggr) = +\infty.
    \end{align}
    Note that the division by $g_{1}(t)$ on the left-hand side of the above expression is allowed for $t>0$ (large enough) because by the definition of $g_{1}(t)$, once an asymptotically stable subsystem is activated, $g_{1}(t)>0$. By the standard properties of $\li$, $\grave{a}$ la \cite[\S0.1]{LojaReal},
    \begin{align}
    \label{e:proof5}
        \li_{t\to+\infty}\Biggl(g_{1}(t)\biggl(1-\frac{g_{2}(t)}{g_{1}(t)}\biggr)\Biggr) \ge \li_{t\to+\infty}g_{1}(t)\li_{t\to+\infty}\biggl(1-\frac{g_{2}(t)}{g_{1}(t)}\biggr),
    \end{align}
    because the right-hand side is not of the form $0\cdot\pm\infty$. Also, by definition of $g_{1}(t)$,
    \[
        \li_{t\to+\infty}g_{1}(t) \in\: ]0,+\infty].
    \]
    Thus, to ensure that the right-hand side of \eqref{e:proof5} diverges, we need to find conditions such that
    \[
        \li_{t\to+\infty}\biggl(1-\frac{g_{2}(t)}{g_{1}(t)}\biggr) > 0,
    \]
    which by standard properties of $\li$ $\grave{a}$ la \cite[\S0.1]{LojaReal}, is equivalent to
    \[
        \ls_{t\to+\infty}\biggl(\frac{g_{2}(t)}{g_{1}(t)}-1\biggr) < 0,
    \]
    which is equivalent to
    \begin{align}
    \label{e:proof6}
        \ls_{t\to+\infty}\frac{g_{2}(t)}{g_{1}(t)} < 1.
    \end{align}
    Employing the definitions of $g_{1}(t)$ and $g_{2}(t)$ from \eqref{e:g1defn} and \eqref{e:g2defn}, respectively, in \eqref{e:proof6}, we see that \eqref{e:thmcondn} holds.

    It remains to verify \eqref{e:gascondn2}. To this end, we get back to \eqref{e:proof4}. Since by \eqref{e:gascondn1}, $\exp{\bigl(g_{2}(t)-g_{1}(t)\bigr)}$ is convergent, it is bounded. Let $\displaystyle{[0,+\infty[\:\ni r\longmapsto \overline{\alpha}(r) \Let \lambda_{\max}\biggl(\sum_{i\in\P}P_{i}\biggr)r^{2}}$ and $\displaystyle{[0,+\infty[\:\ni r\longmapsto \underline{\alpha}(r) \Let \min_{i\in\P}\lambda_{\min}(P_{i})r^{2}}$, we see that
    \[
        \underline{\alpha}(\norm{z}) \le V_{i}(z) \le \overline{\alpha}(\norm{z})\:\:\text{for all}\:\:i\in\P\:\:\text{and}\:\:z\in\R^{d}.
    \]
    In conjunction with \eqref{e:proof4}, we get
    \begin{align*}
        \underline{\alpha}(\norm{z(t)}) \le V_{\sigma(t)}(x(t)) \le \overline{\alpha}(\norm{x_{0}})\exp{\bigl(g_{2}(t)-g_{1}(t)\bigr)}\\
        \text{for all}\:\:t\in\N_{0},
    \end{align*}
    which implies, for $\displaystyle{c = \sqrt{\frac{\lambda_{\max}\bigl(\sum_{i\in\P}P_{i}\bigr)}{\min_{i\in\P}\lambda_{\min}(P_{i})}}}$,
    \begin{align}
    \label{e:proof7}
        \norm{x(t)} \le c\norm{x_{0}}\exp{\bigl(g_{2}(t)-g_{1}(t)\bigr)}\:\:\text{for all}\:\:t\in\N_{0}.
    \end{align}
    Since on the right-hand side of \eqref{e:proof7}, the initial condition $x_{0}$ is decoupled from $g_{2}(t)-g_{1}(t)$ and $g_{2}(t)-g_{1}(t)$ depends on $\sigma$, if $\norm{x(t)}<\epsilon$ for all $t>T(\norm{x_{0}},\epsilon)$ for some pre-assigned $\epsilon > 0$, then the solution $(\tilde{x}(t))_{t\in\N_{0}}$ to \eqref{e:swsys} corresponding to an initial condition $\tilde{x}_{0}$ such that $\norm{\tilde{x}_{0}} \le \norm{x_{0}}$ satisfies $\norm{\tilde{x}(t)} < \epsilon$ for all $t > T(\norm{x_{0}},\epsilon)$. The assertion of Theorem \ref{t:mainres} follows at once.
\end{proof}

\begin{proof}[of theorem \ref{t:graphres}]
    \renewcommand{\theenumii}{\alph{enumii}}
        \ref{t2:stepa} Consider a closed walk $W$ on \(G(\P, E(\P))\) of length $\tau$ that satisfies
        \begin{align}
        \label{e:proof2ratio}
            \frac{N(W)}{D(W)} < 1,
        \end{align}
        where
        \begin{align}
        \label{e:Ndefn}
            N(W) \Let \sum_{(k,\ell)\in E(\P)}(\ln\mu_{k\ell})\rho_{k\ell}(W) + \sum_{j\in\P_{U}}\abs{\ln\lambda_{j}}\kappa_{j}(W),
        \end{align}
        and
        \begin{align}
        \label{e:Ddefn}
            D(W) \Let \sum_{j\in\P_{AS}}\abs{\ln\lambda_{j}}\kappa_{j}(W).
        \end{align}
        Fix $t \ge \tau$. Let a walk $W'$ of length $t$ be constructed by repeating the closed walk $W$ of length $\tau$. Therefore,
        \begin{align}
        \label{e:proof21}
            \frac{N(W')}{D(W')} = \frac{\lfloor\frac{t}{\tau}\rfloor N(W)+N(W'')}{\lfloor\frac{t}{\tau}\rfloor D(W)+D(W'')}
        \end{align}
        where $W''$ is a walk of length $t-\lfloor\frac{t}{\tau}\rfloor\tau$. As $t\to+\infty$, $N(W'')$ and $D(W'')$ are negligible compared to $N(W')$ and $D(W')$, respectively in the sense that $N(W'') = o(\lfloor\frac{t}{\tau}\rfloor)$ and $D(W'') = o(\lfloor\frac{t}{\tau}\rfloor)$ as $t \to +\infty$. Consequently,
        \begin{align*}
            \ls_{t\to+\infty}\frac{N(W')}{D(W')} = \frac{\lfloor\frac{t}{\tau}\rfloor N(W)}{\lfloor\frac{t}{\tau}\rfloor D(W)} = \frac{N(W)}{D(W)} < 1 \quad\text{by \eqref{e:proof2ratio}},
        \end{align*}
        and the assertion in Theorem \ref{t:graphres}(a) follows.

        \ref{t2:stepb} \textsf{Step 1:} We first show that every feasible solution to \eqref{e:optprob} is a trail. Recall from \S\ref{s:mainres} the definition of (node arc) incidence matrix $A$ of the directed graph $G(\P,E(\P))$. By the Corollary to Theorem 13.3 \cite[p.\ 318]{papa_optimization}, the feasibility problem \eqref{e:optprob} has only integer optimal solutions. By the constraint $0\le f_{j} \le 1$ for all $1 \le j \le \abs{E(\P)}$, each element of the vector $f$ is either zero or one. Consequently, the solution to \eqref{e:optprob} (if any) is a trail.

        \textsf{Step 2:} It remains to verify that every feasible solution to \eqref{e:optprob} is a circuit. Suppose there exists a feasible solution which is a trail but not a circuit. By definition, the trail begins at a vertex $u\in\P$ and ends at a vertex $v\in\P\backslash\{u\}$. Then, $a_{u}f = +1$ and $a_{v}f = -1$, where $a_{u}$ and $a_{v}$ denote the rows of $A$ corresponding to the distinct vertices $u$ and $v$, respectively. Consequently, the vector $Af$ has $+1$ and $-1$ values for the $u$-th and $v$-th row, respectively, and that contradicts our constraint that $Af = (0,\cdots,0)^\top$ for every feasible \(f\). It follows at once that every solution to \eqref{e:optprob} is a circuit.

        \ref{t2:stepc} Let $G'(\P',E(\P'))$ denote a subgraph of the directed graph $G(\P,E(\P))$ such that the set of vertices $\P'$ and the set of edges $E(\P')$ contain the elements of $\P$ and $E(\P)$ which are included in the circuit that satisfies condition \eqref{e:graphcondn}, respectively. It is possible to construct $G'(\P',E(\P'))$ from the solution $f$ to the feasibility problem \eqref{e:optprob} (if the solution exists).
            Since $f$ represents a circuit (a closed walk with distinct edges), every vertex in $G'(\P',E(\P'))$ has even degree, i.e., $G'(\P',E(\P'))$ is Eulerian \cite[p.56]{Harris}.
            Given the Eulerian graph $G'(\P',E(\P'))$, we apply Hierholzer's algorithm \cite[p.57]{Harris} to obtain an Eulerian circuit on $G'(\P',E(\P'))$.
\end{proof}
%========================================================================

%========================================================================
\section*{Acknowledgement}
%======================================================================
    We thank Ankur Kulkarni for discussions and a pointer to reference \cite{papa_optimization}. We also thank Niranjan Balachandran for several helpful discussions.
%==========================================================================

%============================================================================

%=============================================================================
\end{document}